\documentclass[twocolumn,showpacs,preprintnumbers,amsmath,amssymb]{revtex4}
\usepackage{graphicx}
\usepackage{bm}
\usepackage{dcolumn}

\def\prl#1#2#3{{ Phys.   Rev.   Lett.  } {\bf #1}, #2 (#3)}

\def\pra#1#2#3{Phys.   Rev.   A {\bf #1}, #2 (#3)}
\def\svp#1#2#3{Sov. Phys.-JETP {\bf #1}, #2 (#3)}

\def\epl#1#2#3{{Europhys. Lett.  } {\bf #1}, #2 (#3)}
\def\pre#1#2#3{Phys.   Rev.   E {\bf #1}, #2 (#3)}
\def\pknaw#1#2#3{Prok. Kon. N. Akad. Wet {\bf #1}, #2 (#3)}
\def\njp#1#2#3{{New Journal of Physics  } {\bf #1}, #2 (#3)}
\def\ps#1#2#3{{Physica  } {\bf #1}, #2 (#3)}

\def\noi{\noindent}
\def\bc{\begin{center}}
\def\ec{\end{center}}
 \newcommand{\bea}{\begin{equation}}
 \newcommand{\eea}{\end{equation}\noi}
 \newcommand{\ber}{\begin{eqnarray}}
 \newcommand{\eer}{\end{eqnarray}\noi}
\begin{document}
\title{Bose-Einstein Condensation and Casimir Effect of Trapped Ideal Bose Gas in between two Slabs}
\author{Shyamal Biswas}\email{tpsb@mahendra.iacs.res.in} 
\affiliation{Department of Theoretical Physics \\
Indian Association for the Cultivation of Science \\
Jadavpur,Kolkata-700032, India}

\date{\today}
\begin{abstract}

               We study the Bose-Einstein condensation for a 3-d system of ideal Bose gas which is harmonically trapped along two perpendicular directions and is confined in between two slabs along the other perpendicular direction. We calculate the Casimir force between the two slabs for this system of trapped Bose gas. At finite temperatures this force for thermalized photons in between two plates has a classical expression which is independent of $\hbar$. At finite temperatures the Casimir force for our system depends on $\hbar$. For the calculation of Casimir force we consider only the Dirichlet boundary condition. We show that below condensation temperature($T_c$) the Casimir force for this non-interacting system decreases with temperature($T$) and at $T\gtrsim T_c$, it is independent of temperature. We also discuss the Casimir effect on 3-d highly anisotropic harmonically trapped ideal Bose gas.      
\end{abstract}

\maketitle
           Vacuum fluctuation of electromagnetic field would cause an attractive force between two closely spaced parallel conducting plates. This phenomenon is called Casimir effect and this force is called Casimir force \cite{1}\cite{2}\cite{3}. In the original paper\cite{1} the Casimir force at zero temperature ($T=0$) was defined as 
\bea
F_c(L)=-\frac{\partial}{\partial L}[E(L)-E(\infty)]
\eea
where $E(L)$ is the ground state energy (i.e. the vacuum energy) of the electromagnetic field in between the two conducting plates separated at a distance $L$. However, it is generalized \cite{4} for any range of temperature and for any dielectric substance in between two dielectric plates. Casimir Effect is also generalized for thermodynamical systems \cite{5}. At finite temperature $T$, the definition of Casimir force is generalized as \cite{6}\cite{7}
\bea
F_c(T,L)=-\frac{\partial}{\partial L}[\Omega_T(L)-\Omega_T(\infty)] 
\eea
where $\Omega_T(L)$ is the grand potential of the system in between the plates separated at a distance $L$.

           We consider the Casimir effect for thermodynamical system in particular for the case of Bose gas in between two slabs. Geometry of the system on which some external boundary condition can be imposed is responsible for Casimir effect. Thermalized photons (massless bosons) in between two conducting plates of area $A$ at temperature $T$ gives rise to the Casimir pressure as \cite{8}\cite{9}\cite{10}
\begin{eqnarray}
\frac{F_c(L)}{A}\sim &-&\frac{\pi^{2}\hbar c}{240L^{4}}[1+\frac{16(k_BT)^{4}L^{4}}{3(\hbar c)^{4}}]\   \ for\  \frac{\pi\hbar c}{k_BTL}\gg 1 \nonumber\\ \sim &-&\frac{k_BT\zeta(3)}{8\pi L^{3}}\    \ for\  \frac{\pi\hbar c}{k_BTL}\rightarrow 0
\end{eqnarray}
where $k_B$ is the Boltzmann constant, $c$ is the velocity of light and $L$ is the separation of the parallel plates. At $T\rightarrow 0$, Casimir pressure becomes $-\frac{\pi^{2}\hbar c}{240L^{4}}$ and it is only the vacuum fluctuation which contributes to the Casimir pressure. At high temperature i.e. for $\frac{\pi\hbar c}{k_BTL}\rightarrow 0$, the Casimir force for photon gas  goes as $L^{-3}$ and has a purely classical expression independent of $\hbar$. Casimir effect for ideal massive Bose gas has been studied in \cite{6}. It has been shown that below the condensation temperature $T_c$, Casimir force decreases with temperature \cite{6}.    

       Let us consider a system of trapped Bose gas in between two slabs of width $L$ and consider the bosons of mass $m$ to be oscillating with the same angular frequency $(\omega)$ along the two perpendicular openings of the slabs. Let us also consider that the system is in thermodynamic equilibrium with its surroundings at temperature $T$. The system behaves as isotropic trapped harmonic oscillators along two perpendicular directions and along the other perpendicular direction it behaves like particles is 1-d box. At thermodynamic limit the system size must be greater greater than the thermal de Broglie wavelength $\sqrt{\frac{\pi\hbar^{2}}{2mk_BT}}$ of the particles. For a slab geometry of a 3-d system the length scale along one direction is much less than that along the other two perpendicular directions. For this system of Bose gas we shall show that the Casimir force will depend on $\hbar$.

        We define the thermodynamic limit as the total no. of particles $N\rightarrow\infty$, $ \frac{L}{\sqrt{\frac{\pi\hbar^{2}}{2mk_BT}}}\rightarrow\infty$ and $\frac{k_BT}{\hbar\omega}\rightarrow\infty$ such that $\frac{N\omega^{2}}{L}=constant$. Energy of a single particle state \\$\mid n,j\rangle$ of this system of ideal Bose gas is $E_{n,j}=(n+1)\hbar\omega+\frac{\pi^{2}\hbar^{2}}{2mL^{2}}$ where $n=0,1,2...$ and $j=1,2,3...$. Since the trapped frequencies are the same for the two perpendicular directions the single particle energy state $\mid n,j\rangle$ has degeneracy $(n+1)$. At the thermodynamic limit we can write the single particle energy as $E_{n,p}=n\hbar\omega +\frac{p^{2}}{2m}$ where $p$ is the momentum along the direction in which the particles are free. At this limit we can also write the degeneracy $(n+1)$ as $n$. Considering the thermodynamic limit the total number of thermally excited particles is
\bea
N_T=\int_{-\infty}^{\infty}\int_{0}^{\infty}n\frac{1}{e^{[\frac{p^{2}}{2m}+n\hbar\omega-\mu]/k_BT}-1}dn\frac{Ldp}{2\pi\hbar}
\eea
where $\mu$ is the chemical potential. Here the pre-factor($n$) in the integrand is the degeneracy. Bose condensation temperature ($T_c$) is defined as a temperature where all the particles are thermally excited and below that temperature a macroscopic number of particles come to the ground state \cite{11}\cite{12}\cite{13}. At $T\le T_c$ the chemical potential goes to the ground state energy. So
\begin{eqnarray}
N&=&\int_{-\infty}^{\infty}\int_{0}^{\infty}n\frac{1}{e^{[\frac{p^{2}}{2m}+n\hbar\omega]/k_BT_c}-1}dn\frac{Ldp}{2\pi\hbar}\nonumber\\ &=&\sum_{i=1}^{\infty}\int_{-\infty}^{\infty}\int_{0}^{\infty}ne^{\frac{-i[p^{2}]}{2mk_BT_c}} e^{\frac{-i[n\hbar\omega]}{k_BT_c}}dn\frac{Ldp}{2\pi\hbar}\nonumber\\&=&(\frac{k_BT_c}{\hbar\omega})^{2}\frac{1}{2}\frac{L}{\lambda_c}\zeta(5/2)
\end{eqnarray}
where $\lambda_c =\sqrt{\frac{\pi\hbar^{2}}{2mk_BT_c}}$. So the condensation temperature is 
\bea
T_c=\frac{1}{k_B}[\frac{2\pi\hbar^{6}\omega^{4}}{m L^{2}}]^{\frac{1}{5}}N^{\frac{2}{5}}
\eea

          However for a finite system the total no. of particles are not going to infinity. The angular frequency $\omega$ also does not go to zero. The length $L$ also does not go to infinity. For finite system we consider the thermodynamic limit as  $N \gg 1$, $ L \gg \sqrt{\frac{\pi\hbar^{2}}{2mk_BT}}$ and $\frac{k_BT}{\hbar\omega}\gg 1$ such that $\frac{N\omega^{2}}{L}=constant$. For a slab geometry of a 3-d system the length scale along one direction is much less than that along other two perpendicular directions. Similarly the condition of slab geometry for our system is
\bea
\frac{\pi^{2}\hbar^{2}}{2mL^{2}}\gg  \hbar\omega 
\eea
We take $\frac{\hbar\omega}{k_BT}\rightarrow 0$. The ground state energy of our system is $[g=\hbar\omega +\frac{\pi^{2} \hbar^{2}}{2mL^{2}}]$. Now the average no. of particles with energy $E_{n,j}$ is given by $\frac{1}{e^{[n\hbar\omega +\frac{\pi^{2} \hbar^{2}(j^{2}-1)}{2mL^{2}}+\mu']/k_BT}-1}$ where $\mu'=g-\mu\ge 0$ for bosons. At and below the condensate temperature $\mu'\rightarrow 0$. So for our system of trapped Bose gas the grand potential is $\Omega=\Omega(\omega,L,T,\mu')$. For this bosonic system we have the grand potential as
\begin{eqnarray}
&&\Omega=\Omega(\omega,L,T,\mu')\nonumber\\
&&=k_BT\sum_{n=0}^{\infty}\sum_{j=1}^{\infty}(n+1)\log [1-e^{\frac{-(n\hbar\omega +\frac{\pi^{2}\hbar^{2}(j^{2}-1)}{2m L^{2}}+\mu')}{k_BT}}]\nonumber\\
\end{eqnarray}
The pre-factor $(n+1)$ of the above eqn.(8) is the degeneracy of the single particle state. The number $1$ of this term $(n+1)$  contribute insignificantly to the Casimir force. So we write $n$ instead of $(n+1)$ in the grand potential. In eqn.(8) we also replace $j$ by $(j'+1)$. So from the above equation (8) we can write
\begin{eqnarray}
&&\Omega(\omega,L,T,\mu')\nonumber\\&&\approx-k_BT\sum_{n=0}^{\infty}\sum_{j'=0}^{\infty}\sum_{i=1}^{\infty}n\frac{e^{-\frac{i\mu'}{k_BT}}e^{-\frac{ni\hbar\omega}{k_BT}}e^{-i(\pi(\frac{\lambda}{L})^{2}[j'^{2}+2j'])}}{i}\nonumber\\
\end{eqnarray}
where $\lambda =\sqrt{\frac{\pi\hbar^{2}}{2mk_BT}}$.
Since $\frac{\hbar\omega}{k_BT}\rightarrow 0$ , converting the summation over n into the integration we can write
\begin{eqnarray}
&&\Omega(\omega,L,T,\mu')\nonumber\\&&=-k_BT[\frac{k_BT}{\hbar\omega}]^{2}\sum_{i=1}^{\infty}\sum_{j'=0}^{\infty}\frac{e^{-i\mu'/k_BT}}{i^{3}}[e^{\frac{-\pi i \lambda^{2}(j'^{2}+2j')}{L^{2}}}]
\nonumber\\&&=-k_BT[\frac{k_BT}{\hbar\omega}]^{2}\sum_{i=1}^{\infty}\sum_{j'=0}^{\infty}\frac{e^{-i\mu'/k_BT}}{i^{3}}[e^{\frac{-\pi i \lambda^{2}j'^{2}}{L^{2}}}][1\nonumber\\&&-2j'\frac{\pi i \lambda^{2}}{L^{2}}+2j'^{2}(\frac{\pi i \lambda^{2}}{L^{2}})^{2}-\frac{4}{3}j'^{3}(\frac{\pi i \lambda^{2}}{L^{2}})^{3}+...]
\end{eqnarray}
Since $\frac{\lambda}{L}\ll 1$, higher order terms of the above series would contribute insignificantly. From Euler-Maclaurin summation formula we convert the summation over $j'$ to integration. So from equation(10) we have
\begin{eqnarray}
&&\Omega(\omega,L,T,\mu')\nonumber\\&&=-k_BT[\frac{k_BT}{\hbar\omega}]^{2}\sum_{i=1}^{\infty}\frac{e^{-i\mu'/k_BT}}{i^{3}}[(\int_{0}^{\infty}e^{\frac{-\pi i \lambda^{2}j'^{2}}{L^{2}}}dj'+\frac{1}{2})-\nonumber\\&&2\frac{\pi i \lambda^{2}}{L^{2}}(\int_{0}^{\infty}j'e^{\frac{-\pi i \lambda^{2}j'^{2}}{L^{2}}}dj'-\frac{1}{12})+2[\frac{\pi i \lambda^{2}}{L^{2}}]^{2}\nonumber\\&&(\int_{0}^{\infty}j'^{2}e^{\frac{-\pi i \lambda^{2}j'^{2}}{L^{2}}}dj')-\frac{4}{3}[\frac{\pi i \lambda^{2}}{L^{2}}]^{3}(\int_{0}^{\infty}j'^{3}e^{\frac{-\pi i \lambda^{2}j'^{2}}{L^{2}}}dj'\nonumber\\&&+\frac{6}{720})+...] 
\end{eqnarray}
From the above equation(11) we have
\begin{eqnarray}
&&\Omega(\omega,L,T,\mu')\nonumber\\&&=-k_BT[\frac{k_BT}{\hbar\omega}]^{2}\sum_{i=1}^{\infty}\frac{e^{-i\mu'/k_BT}}{i^{3}}[\frac{L}{2\lambda i^{1/2}}-\frac{1}{2}+\nonumber\\&&\frac{\pi}{2}i^{1/2}\frac{\lambda}{L}+\it{O}([\frac{\lambda}{L}]^{2})]
\end{eqnarray}
        Let us now calculate the Casimir force. At $T\le T_c$ we put $\mu'\rightarrow 0$. So from eqn.(12) we have
\begin{eqnarray}
\Omega(\omega,L,T,0)
 =&-&k_BT[\frac{k_BT}{\hbar\omega}]^{2}[\frac{L}{2\lambda}\zeta(7/2)-\frac{1}{2}\zeta(3)\nonumber\\&&
+\frac{\pi}{2}\zeta(5/2)\frac{\lambda}{L}]
\end{eqnarray}

 Here the first term of eqn. (13) is
\bea
\Omega_b=-k_BT[\frac{k_BT}{\hbar\omega}]^{2}[\frac{L}{2\lambda}\zeta(7/2)].
\eea
It is the bulk term of the grand potential. From our consideration of thermodynamic limit $\frac{N\omega^{2}}{L}=constant$. So $\Omega_T(\infty)=\Omega_b$. The second term of eqn. (13) is $(\Omega_s)=k_BT[\frac{k_BT}{\hbar\omega}]^{2}[\frac{1}{2}\zeta(3)]$. It is the surface term of the grand potential.  The third term of eqn.(13) is the Casimir term of the grand potential. We say it Casimir potential. Now putting $\lambda =\sqrt{\frac{\pi\hbar^{2}}{2mk_BT}}$ in eqn.(13) we have the Casimir potential as
\bea
\Omega_c=-k_BT[\frac{k_BT}{\hbar\omega}]^{2}\sqrt{\frac{\pi\hbar^2}{2mL^{2}k_BT}}\frac{\pi}{2}\zeta(5/2) 
\eea
From eqn. (15),(5) and (6) we can write
\bea
\Omega_c=-N[\frac{T}{T_c}]^{5/2}\frac{\pi^{2}\hbar^{2}}{2mL^{2}}
\eea
So at $T\le T_c$, from eqn.(2),(13),(14) and (16) we have the expression of Casimir force as
\bea
F_c(T,L)=-N[\frac{T}{T_c}]^{5/2}\frac{\pi^{2}\hbar^{2}}{mL^{3}}\   \ for \ T\le T_c
\eea
Finally from eqn.(18) we have a macroscopic Casimir force. The expressions of Casimir force in eqn.(18) shows that, the finite temperature Casimir force for this particular system depends on $\hbar$. These expressions are no longer classical.         

          Above the condensation temperature, $\mu'>0$. At $T\gtrsim T_c$, the expression of total number of particles, instead of eqn.(5) can be written as 
\bea
N=(\frac{k_BT}{\hbar\omega})^{2}\frac{1}{2}\frac{L}{\lambda}g_{5/2}(e^{-\mu'/k_BT})
\eea
where and $g_{5/2}(e^{-\mu'/k_BT})$ is Bose-Einstein condensation function which is defined as 
$g_{5/2}(x)=x+\frac{x^2}{2^{5/2}} +\frac{x^3}{3^{5/2}}+\frac{x^4}{4^{5/2}}+....$

 So at $T\gtrsim T_c$, from the eqn.(12), with trivial manipulation we get the Casimir potential as 
\begin{eqnarray}
\Omega_c&=&-k_BT[\frac{k_BT}{\hbar\omega}]^{2} \sum_{i=1}^{\infty}\frac{e^{-i\mu'/k_BT}}{i^{3}}[\frac{\pi}{2}i^{1/2}\frac{\lambda}{L}]\nonumber\\&=&-k_BT[\frac{k_BT}{\hbar\omega}]^{2}\frac{\pi\lambda}{2L}g_{5/2}(e^{-\mu'/k_BT})\nonumber\\&=& -\pi N k_B T (\frac{\lambda}{L})^{2}\nonumber\\&=&-N\frac{\pi^{2}\hbar^{2}}{2mL^{2}} 
\end{eqnarray}
So, at $T\gtrsim T_c$, from the definition of Casimir force we have 
\bea
F_c(T,L)=-N\frac{\pi^{2}\hbar^{2}}{mL^{3}}\   \ for \ T\gtrsim T_c
\eea 
So at temperatures $T\gtrsim T_c$, the Casimir force is independent of temperature.
           
           In case of 3-d harmonically trapped Bose gas whether Bose-Einstein condensation has been achieved is experimentally determined from the speed distribution of the atoms. But, for this system the appearance of Casimir force at some temperature($T_c$) and the reduction of this force  below this temperature with $T^{5/2}$ law may signalize the occurrence of BEC. With $10^{9}$ hydrogen atoms Bose-Einstein condensation has been performed \cite{14}. Typical Casimir force of the order of $10^{-12}$N with $L\sim 10^{-7}$m has been experimentally performed \cite{15}. For the experimental set up if we choose $\omega\sim 100s^{-1}$,$L\sim10^{-7}$m.,$N\sim10^{9}$ then the slab condition(7) will be satisfied for light alkali atoms. These values of parameters will give $T_c\sim10^{-5}$K and $F_c(T,L)\sim10^{-12}$ Newton for alkali atoms.
   
            However, we can think of the Casimir effect for 3-d harmonically anisotropic trapped Bose gas. We take very asymmetric harmonic trap with two different frequencies($\omega,\omega_o$) such that $\omega_o\gg\omega$. For this system single particle energy is $E(n,n_o)=(n+1)\hbar\omega+(n_o+\frac{1}{2})\hbar\omega_o$ where $n,n_o=0,1,2...$ For this type of trap geometry we write the grand potential as 
\begin{eqnarray}
&&\Omega=\Omega(\omega,\omega_o,T,\mu')\nonumber\\
&&=k_BT\sum_{n=0}^{\infty}\sum_{n_o=0}^{\infty}(n+1)\log [1-e^{\frac{-(n\hbar\omega +n_o\hbar\omega_o+\mu')}{k_BT}}]\nonumber\\
\end{eqnarray}
Here $\mu'= (\hbar\omega +\frac{1}{2}\hbar\omega_o -\mu)$. At and below the condensation temperature $\mu'=0$. Putting $\mu'=0$ in the above eqn.(21) and from the similar type of calculations as shown above we can write the grand potential as
\bea
\Omega=-k_BT\sum_{i=1}^{\infty}[(\frac{k_BT}{\hbar\omega})^{2}\frac{1}{i^{3}}+(\frac{k_BT}{\hbar\omega})\frac{1}{i^{2}}]\frac{1}{(1-e^{-\frac{i\hbar\omega_o}{k_BT}})}
\eea
Neglecting the contribution of the second term of the square bracket of the above equation and expanding the exponential term into series and with a trivial manipulation we get
\begin{eqnarray}
\Omega=&-&k_BT(\frac{k_BT}{\hbar\omega})^{2}[(\frac{k_BT}{\hbar\omega_o})\zeta(4)+\frac{1}{2}\zeta(3)\nonumber\\&+&\frac{1}{12}(\frac{\hbar\omega_o}{k_BT})\zeta(2)+\it{O}([\frac{\hbar\omega_o}{k_BT}]^2)]    
\end{eqnarray} 
The expression of the total number of particles for this system is\cite{12} $N=(\frac{k_BT_c}{\hbar\omega})^{2} \frac{k_BT_c}{\hbar\omega_o}\zeta(3)$.           
The third term of the above eqn.(23) is the Casimir potential($\Omega_c$). For this geometry we also see that the Casimir potential $\Omega_c=-\frac{1}{12}(\frac{k_BT}{\hbar\omega})^{2}\zeta(2)\hbar\omega_o=N\frac{\hbar\omega_o\zeta(2)}{12k_BT_c\zeta(3)}[\frac{T}{T_c}]^{2}\hbar\omega_o$. For this system of 3-d trapped geometry, below $T_c$, the Casimir potential is quadratic in temperature.  
         
            Due to the symmetry of the wave function an `effective attracting force' sets in between the ideal Bose particles of different quantum mechanical states. If the system size is finite then this `effective attracting force' considerably reduces the pressure of the system. In other words this force gives rise to thermodynamic Casimir force. At $T< T_c$, a macroscopic no. particles come to the ground state. There is no `effective attracting force' for the particles in a single particle state due to wave function symmetry. So, at $T< T_c$, the macroscopic no. of particles in the ground state would not contribute to the Casimir force. Due to this fact, below $T_c$, Casimir force decreases as we decrease the temperature. At $T\gtrsim T_c$, the Casimir force for this non-interacting system is independent of temperature for the slab geometry. But, at $T\gtrsim T_c$, the Casimir potential decreases with temperature for anisotropic geometry of ideal trapped Bose gas. However at very high temperature ($T\gg T_c$) and in the classical limit, Bose-Einstein statistics will be ineffective and the Casimir Force will vanish. Below $T_c$ the thermodynamic Casimir force reduction is the signature of Bose-Einstein condensation. 
          
           Several useful discussions with Jayanta Kumar\\ Bhattacharjee of I.A.C.S. and with Debnarayan Jana of C.U. are gratefully acknowledged.

\end{document}